# Elimination of a Second-Law-attack, and all cable-resistance-based attacks, in the Kirchhoff-law–Johnson-noise (KLJN) secure key exchange system

**Laszlo B. Kish [1],*** **and Claes-Göran Granqvist [2]**

[1] Department of Electrical and Computer Engineering, Texas A&M University, College Station, TX 77843-3128, USA

[2] Department of Engineering Sciences, The Ångström Laboratory, Uppsala University, P. O. Box 534, SE-75121 Uppsala, Sweden; E-Mail: claes-goran.granqvist@angstrom.uu.se

The authors contributed equally to this work.

* Author to whom correspondence should be addressed; E-Mail: Laszlo.Kish@ece.tamu.edu

**Abstract:** We introduce the so far most efficient attack against the Kirchhoff-law–Johnson-noise (KLJN) secure key exchange system. This attack utilizes the lack of exact thermal equilibrium in practical applications and is based on cable resistance losses and the fact that the Second Law of Thermodynamics cannot provide full security when such losses are present. The new attack does not challenge the unconditional security of the KLJN scheme, but it puts more stringent demands on the security/privacy enhancing protocol than for any earlier attack. In this paper we present a simple defense protocol to fully eliminate this new attack by increasing the noise-temperature at the side of the smaller resistance value over the noise-temperature at the at the side with the greater resistance. It is shown that this simple protocol totally removes Eve's information not only for the new attack but also for the old Bergou-Scheuer-Yariv attack. The presently most efficient attacks against the KLJN scheme are thereby completely nullified.

## 1. Introduction

The Kirchhoff-law–Johnson-noise (KLJN) scheme [1,2], shown in Figure 1, is a classical statistical physical competitor to a quantum key distribution for secure communication. For the duration of a single bit exchange, the communicating parties (Alice and Bob) connect their randomly chosen resistor and corresponding noise-voltage generator to a wire line (cable). These resistors are randomly selected from the publicly known set $\{R_L, R_H\}$, $R_L \neq R_H$, where the elements represent low (L) and high (H) bit values. The Gaussian voltage noise generators—mimicking the Fluctuation-Dissipation Theorem and delivering band-limited white noise with publicly agreed bandwidth—produce enhanced thermal (Johnson) noise at a publicly agreed effective temperature $T_{eff}$, typically being $T_{eff} \geq 10^9 \text{K}$ [3], so the temperature of the wire can be neglected. The noises are statistically independent of each other and from the noise of the former bit period.

In the case of secure bit exchange—*i.e.*, the LH or HL bit situations for Alice and Bob—an eavesdropper (Eve) cannot distinguish between these two situations by measuring the mean-square value of the voltage $U_c(t)$ and/or current $I_c(t)$ in the cable, because both arrangements lead to the same result. In the rest of the paper we assume that one of these secure bit exchange situations (either LH or HL) apply.

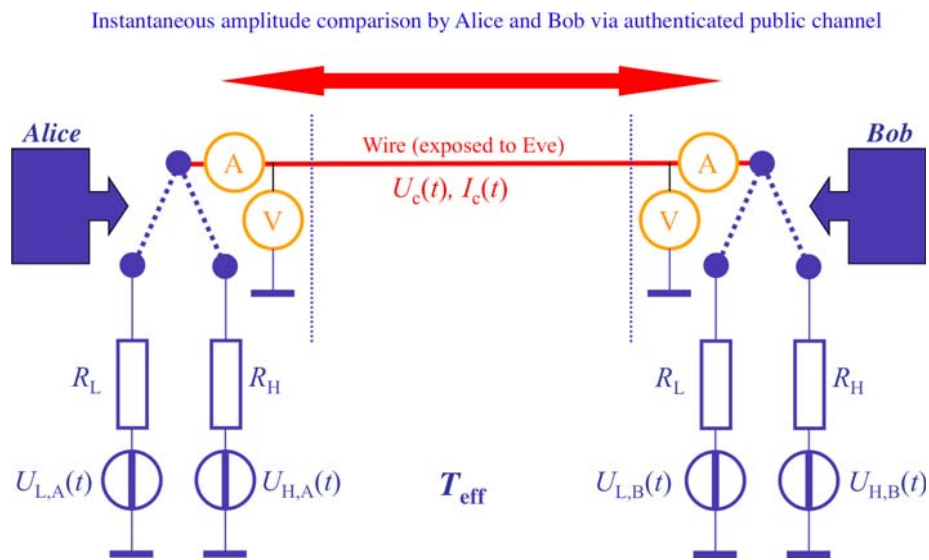

**Figure 1.** Schematic of the Kirchhoff-law–Johnson-noise secure key exchange system. To defend against active and hacking attacks, the cable parameters and integrity are randomly monitored; the instantaneous voltage $U_c(t)$ and current $I_c(t)$ amplitudes in the cable are measured and compared via public authenticated data exchange; and full spectral and statistical analysis/checking is carried out by Alice and Bob. $R$, $t$ and $T_{eff}$ denote resistance, time and effective temperature, respectively. Line filters, etc., are not shown.

To avoid potential information leak by variations in the shape of a probability distribution, the noises are Gaussian [1], and it has been proven that other distributions are not secure [4,5]. From a physics perspective, the security is provided by the Second Law of Thermodynamics because directional information, due to the direction of power flow, does not exist since the mean power flow is zero even though the LH and HL situations have asymmetric resistance arrangements [1]. In other words, the security of the *ideal* KLJN scheme against passive (non-invasive listening/measuring) attacks is as strong as the impossibility to build a perpetual motion machine of the second kind. The security against active (invasive) attacks is—perhaps surprisingly—provided by the robustness of classical physical quantities, which guarantees that these quantities can be monitored (and their integrity with the cable parameters and model can be checked) *continuously* without destroying their values. We observe, in passing, that the situation is totally different for the case of quantum physics.

The most famous and explored, and so far the most effective, attack against the *non-ideal* KLJN scheme is the Bergou-Scheuer-Yariv (BSY) cable resistance attack [6,7] which utilizes the fact that, due to the non-zero cable resistance, the mean-square voltage will be slightly less at the cable end with the smaller resistance value than at the other end with the greater resistance. It should be noted that the results (including their physical units) are wrong in Ref. [7], but a correct evaluation of the BSY effect was carried out later by Kish and Scheuer (KS) [8]. Eve's measured absolute difference between the mean-square voltages $\langle U_{cH}^2(t) \rangle$ and $\langle U_{cL}^2(t) \rangle$ of the "H" and "L" ends (*cf.* Figure 2) is given by [8]

$$\Delta_{KS} = \left| \langle U_{cH}^2(t) \rangle - \langle U_{cL}^2(t) \rangle \right| = 4kT_{eff}\Delta f \left| \frac{R_c^2(R_H - R_L)}{(R_H + R_c + R_L)^2} \right|, \quad (1)$$

where $k$ is Boltzmann's constant, $\Delta f$ is noise bandwidth and $R_c$ is cable resistance. Clearly $\Delta_{KS}$ scales with the square of the cable resistance, *i.e.*, $\Delta_{KS} \propto R_c^2$.

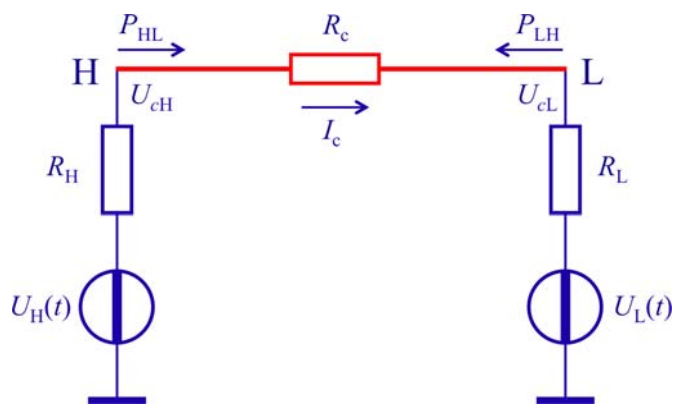

**Figure 2.** Scheme devised to illustrate the Bergou-Scheuer-Yariv attack and the Second-Law-attack. Alice's and Bob's locations are arbitrary in the figure. During the Second-Law-attack, the powers flowing out from the "H" and "L" ends of the cable are calculated and compared. The temperature of the cable resistor $R_c$ can be neglected because of the high noise temperature of the generators. The notation is consistent with that in Figure 1.

## 2. Results and Discussion

### 2.1. The Second-Law-attack

In the rest of the paper we use the rules about transformations of noise spectra in linear systems, along with Johnson's formula for thermal noise, and write [1]

$$\langle U_R^2(t) \rangle = 4kT_{eff} R\Delta f \quad . \quad (2)$$

Here $\langle U_R^2(t) \rangle$ denotes mean-square voltage fluctuations on the resistor, with resistance $R$, within the bandwidth $\Delta f$.

The cable resistance has a non-zero value, and therefore the resistors and their noise generators are not in thermal equilibrium in practical versions of the KLJN system (with $T_{eff}$ much greater than the cable temperature). Consequently the Second Law of Thermodynamics cannot provide full security. The cable-heating powers by the generators at the "H" and "L" ends are different and are given by

$$P_{Hc} = \langle I_A^2(t) \rangle R_c = \frac{4kT_{eff}R_H\Delta f}{(R_H + R_c + R_L)^2} R_c , \quad (3)$$

and

$$P_{Lc} = \langle I_B^2(t) \rangle R_w = \frac{4kT_{eff}R_L\Delta f}{(R_H + R_c + R_L)^2} R_c = P_{Hc}\frac{R_L}{R_H} . \quad (4)$$

The difference between $P_{Hc}$ and $P_{Lc}$ can be utilized for the Second-Law-attack, because the resistor values $R_H$ and $R_L$ are publicly known. The implementation of this attack is to measure and compare the net power flows at the two ends of the cable, as illustrated in Figure 2. The mean power flow $P_{HL}$ from the "H" end toward the "L" end of the cable, and the mean power flow $P_{LH}$ from the "L" end toward the "H" end are, respectively,

$$P_{HL} = \langle U_H^2(t) \rangle \left(\frac{R_c + R_L}{R_H + R_c + R_L}\right)^2 \frac{1}{R_c + R_L}$$
$$- \langle U_L^2(t) \rangle \left(\frac{R_H}{R_H + R_c + R_L}\right)^2 \frac{1}{R_H} \quad (5)$$
$$= 4kT_{eff}\Delta f \frac{R_H(R_c + R_L) - R_L R_H}{(R_H + R_c + R_L)^2} = 4kT_{eff}\Delta f \frac{R_H R_c}{(R_H + R_c + R_L)^2}$$

and

$$P_{LH} = \langle U_L^2(t) \rangle \left(\frac{R_c + R_H}{R_H + R_c + R_L}\right)^2 \frac{1}{R_c + R_H}$$
$$- \langle U_H^2(t) \rangle \left(\frac{R_L}{R_H + R_c + R_L}\right)^2 \frac{1}{R_L} \quad (6)$$
$$= 4kT_{eff}\Delta f \frac{R_L(R_c + R_H) - R_H R_L}{(R_H + R_c + R_L)^2} = 4kT_{eff}\Delta f \frac{R_L R_c}{(R_H + R_c + R_L)^2} .$$

The power flows $P_{HL}$ and $P_{LH}$ are directly measurable by Eve, and their difference,

$$\Delta P_{HL} = P_{HL} - P_{LH} = 4kT_{eff}\Delta f \frac{R_c(R_H + R_L)}{(R_H + R_c + R_L)^2} , \quad (7)$$

gives the difference between the powers supplied by the two cable ends; with the measured cable voltages and current (see Figure 2) it is

$$\Delta P_{HL} = P_{HL} - P_{LH} = \langle I_c(t)U_{cH}(t) \rangle - \langle -I_c(t)U_{cL}(t) \rangle$$
$$= \langle [U_{cH}(t) + U_{cL}(t)]I_c(t) \rangle . \quad (8)$$

It should be observed that the opposite current sign at the "L" end expresses the fact that the current flowing *out* from the "H" end is flowing *into* the "L" end (using the same current sign would instead provide the power dissipated in the cable resistance, which is always positive and gives no directional information).

Suppose now that Eve measures the above current–voltage cross-correlations at the two ends and evaluates the pertinent quantities. With the notation introduced in Figure 3, one finds that

$$\Delta P_{AB} = P_{AB} - P_{BA} = \langle [U_{cA}(t) + U_{cB}(t)] I_c(t) \rangle \quad . \tag{9}$$

As an example, suppose that $R_H$ has the greater resistance value and $R_L$ the smaller one, i.e., $R_L < R_H$. In the *ideal* case, when $R_c = 0$, one obtains $\Delta P_{AB} = 0$ in accordance with the Second Law of Thermodynamics, which yields $\langle U_c(t) I_c(t) \rangle = 0$. However, in the *practical* case, with $R_c > 0$, one finds

(*i*) if $\Delta P_{AB} > 0$, then Alice has $R_H$ and Bob has $R_L$,

(*ii*) if $\Delta P_{AB} < 0$, then Alice has $R_L$ and Bob has $R_H$.

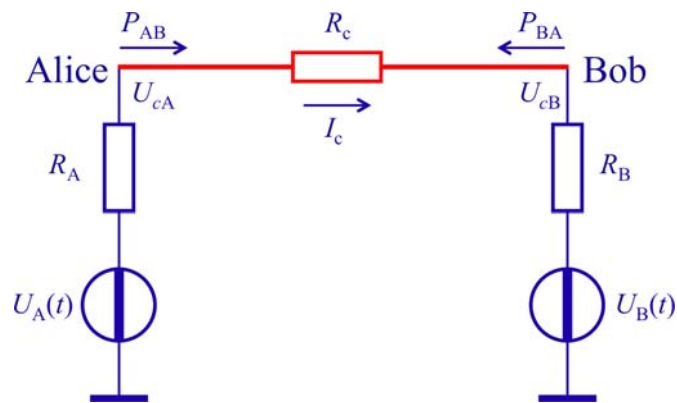

**Figure 3.** Eve's measurements during the Second-Law-attack. The powers flowing out from the two ends of the cable are measured and compared. The notation is consistent with that in Figure 1.

The signal inherent in the Second-Law-attack scales linearly with $R_c$, which provides a much better situation for Eve—especially in the case of vanishing cable resistance—than the square-law scaling of the BSY attack. Moreover, it is also obvious that in a practical case [3,9,10], where $R_c \ll R_L \ll R_H$, Eve's signal-to-noise ratio is always greater in the Second-Law-attack than in the BSY attack. This is so because the BSY attack evaluates the dc fraction of $\approx R_c^2 / (R_L R_H)$ in the measured (empirical) mean-square channel noise voltage, while the Second-Law-attack evaluates the dc fraction of $R_c / R_H$ in the measured mean power flow. It should be noted that the measured mean-square channel noise voltage, and the measured mean power flow, follow similar statistics because they are the time average of the products of Gaussian processes [11].

The Second-Law-attack is an elegant and efficient one, but it *does not challenge the unconditional security of the KLJN scheme* [2]. Eve's probability *p* of successful guessing can arbitrarily approach the limit $p = 0.5$ by proper tuning of the parameters inherent in the KLJN scheme, such as resistances and bandwidth, and privacy amplification can be implemented if needed; this was evaluated in detail elsewhere [2], where relationships were reported between security level, cable parameters and communication speed. Nevertheless the new Second-Law-attack is important and may significantly increase the demands on parameter tuning and/or necessitate elaborate privacy amplification [12], which of course come at a cost.

In the rest of this paper we demonstrate two methods capable of fully eliminating the Second-Law attack. The advanced method nullifies the BSY attack as well.

*2.2. Natural/"Simple" defense*

Suppose it is possible to keep the cable and the resistors at the same temperature. This temperature-equilibration method virtually eliminates any Second-Law-attack information for Eve (but not the information in the BSY-attack, albeit its formula for the information leak is changed).

Temperature equilibration constitutes a very simple defense, but the cable temperature and its possible variations cannot be neglected any longer. If the cable temperature is different from that of the resistors, then the KLJN scheme is vulnerable to the Hao-type attack [13] (see its criticism in [14]). In principle, with cables of homogeneous temperatures, this attack can be avoided if Alice and Bob are able to monitor the temperature value of the cable by resistance and Johnson noise measurements, since they can then choose $T_{\text{eff}}$ to be the same as the cable temperature. While these steps can be taken, the KLJN scheme is no longer simple. Moreover, the mentioned defense method may be unpractical because of the requirement of a homogeneous cable temperature, small noise levels, and since it prohibits the adoption of enhanced KLJN methods wherein Alice and Bob eliminate their own contributions in order to accomplish higher speed, security [9,15] and fidelity [16].

*2.3. Advanced defense, also eliminating all cable resistance attacks*

As we have seen, the cable end with the smaller resistance value emits less power toward the other end, and this is the foundation of the Second-Law-attack. This effect, as well as Eve's related signal, can be completely eliminated by properly changing the *ratio* of the noise-temperatures of the generators for the resistors with the smaller and the greater resistance values (see Figure 4).

Suppose now that we introduce an offset in the noise-temperatures of the generators for the $R_\text{H}$ and the $R_\text{L}$ resistors so that the equation

$$\Delta P_\text{HL} = P_\text{HL}(T_\text{eff}) - P_\text{LH}(\beta T_\text{eff}) = 0 \tag{10}$$

holds, where $T_\text{eff}$ is the noise temperature at the $R_\text{H}$ resistors and $\beta T_\text{eff}$ is the noise temperature of the $R_\text{L}$ resistors. The solution of equation (10) is

$$\beta = \frac{1 + \dfrac{R_\text{c}}{R_\text{L}}}{1 + \dfrac{R_\text{c}}{R_\text{H}}} \quad . \tag{11}$$

This value of $\beta$ for the temperature-offset consequently eliminates Eve's opportunity to use the Second-Law-attack. One finds $\beta > 1$ for $R_\text{L} < R_\text{H}$ and $\beta < 1$ for $R_\text{H} < R_\text{L}$.

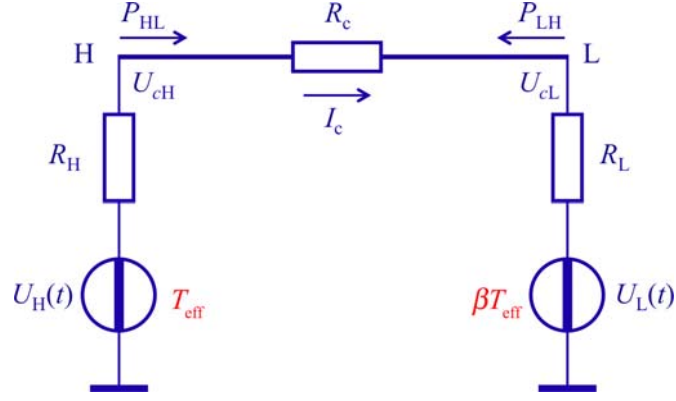

**Figure 4.** Schematic for illustrating the elimination of the Second-Law-attack and the BSY-attack by introduction of a proper temperature offset. The notation is consistent with that in Figure 1.

The remaining, essential question is whether the defense method delineated above introduces a higher signal for Eve's BSY-attack or not. Reevaluating our analysis [8] of the BSY attack with the a temperature offset given by Eq. 11, one obtains

$$\Delta_{KS}(T_{eff}, \beta T_{eff}) = \left| \langle U_{cH}^2(t) \rangle - \langle U_{cL}^2(t) \rangle \right|$$
$$= 4kT_{eff}\Delta f R_H \left| \frac{R_c^2(1-\alpha\beta) - \alpha R_H R_c(\beta-1)}{(R_H + R_c + R_L)^2} \right| , \quad (12)$$

where $\alpha = \frac{R_L}{R_H}$. By substituting the above value for $\beta$, the nominator becomes zero so that

$$\Delta_{KS}(T_{eff}, \beta T_{eff}) = \left| \langle U_{cH}^2(t, T_{eff}) \rangle - \langle U_{cL}^2(t, \beta T_{eff}) \rangle \right| = 0 \quad (13)$$

Hence a modification of the noise temperature of the generators supplying the noise of the $R_L$ resistors by the factor $\beta$ yields a complete elimination the strongest attacks against the KLJN key exchange scheme, namely the Second-Law-attack and the BSY-attack [6–8].

## 3. Conclusions

We introduced the so far most efficient attack against the Kirchhoff-law–Johnson-noise (KLJN) secure key exchanger, *i.e.*, the Second-Law-attack. This attack utilizes the lack of exact thermal equilibrium in practical applications involving cables with non-zero resistance and results in more advantageous scaling and signal-to-noise ratio for Eve.

Late-note: It has come to our attention that Gunn-Allison-Abbott [17] has published a paper about a new type of attack against the KLJN system, where the attack produces signal for Eve only at non-zero wire resistance. Based on our manuscript on arxiv, they tested the defense method described above in section 2.3 and found it working even in their system. We agree with this statement of their paper. Note, we extensively analyzed and criticized all their other claims of their preprint in two separate papers or ours [18,19]. Furthermore, they note that the "measurement" of $\alpha$ may be not obvious [17] from a security point of view. This is a misunderstanding by them [17] because the value of $\alpha$ is a public knowledge in the KLJN protocol [1-3].

## References and Notes

1. Kish, L.B. Totally secure classical communication utilizing Johnson (-like) noise and Kirchoff's law. *Phys. Lett. A*, 2006, **352**, 178–182, DOI: 10.1016/j.physleta.2005.11.062. Available online: http://arxiv.org/abs/physics/0509136.
2. Kish, L.B.; Granqvist C.G. On the security of the Kirchhoff-law-Johnson-noise (KLJN) communicator. *Quantum Inf. Process.*, 2014, in press, DOI: 10.1007/s11128-014-0729-7. Available online:http://arxiv.org/abs/1309.4112,
3. Mingesz, R.; Gingl Z.; Kish L.B. Johnson(-like)–Noise–Kirchhoff-loop based secure classical communicator characteristics, for ranges of two to two thousand kilometers, via model-line. *Phys. Lett. A*, 2008, 372, 978–984, DOI: 10.1016/j.physleta.2007.07.086. Available online: http://arxiv.org/abs/physics/0612153
4. Gingl, Z.; Mingesz, R. Noise Properties in the Ideal Kirchhoff-Law-Johnson Noise Secure Communication System. *PLoS ONE*, 2014, 9, e96109, DOI: 10.1371/journal.pone.0096109. Available online: http://www.plosone.org/article/info%3Adoi%2F10.1371%2Fjournal.pone.0096109
5. Mingesz, R.; Vadai, G; Gingl, Z. What kind of noise guarantees security for the Kirchhoff-Loop-Johnson-Noise key exchange? *Fluct. Noise Lett.*, 2014, 13, 1450021, DOI: 10.1142/S0219477514500217. Available online: http://arxiv.org/abs/1405.1196.
6. Bergou, J., interviewed in: Cho A. Simple Noise May Stymie Spies Without Quantum Weirdness. *Science*, 2005, **309** 2148. DOI: 10.1126/science.309.5744.2148b.
7. Scheuer, J.; Yariv A. A classical key-distribution system based on Johnson (like) noise—How secure?. *Phys. Lett. A*, 2006, 359, 737–740, DOI: 10.1016/j.physleta.2006.07.013.
8. Kish, L.B.; Scheuer J. Noise in the wire: The real impact of wire resistance for the Johnson(-like) noise based secure communicator. *Phys. Lett. A*, 2010, 374, 2140–2142, DOI: 10.1016/j.physleta.2010.03.021. Available online: http://arxiv.org/abs/1002.0087.
9. Kish, L.B. Enhanced Secure Key Exchange Systems Based on the Johnson- Noise Scheme. *Metrol. Meas. Syst.*, 2013, 20, 191–204. DOI: 10.2478/mms-2013-0017. Available online: http://www.degruyter.com/view/j/mms.2013.20.issue-2/mms-2013-0017/mms-2013-0017.xml?format=INT
10. Kish, L.B.; Abbott, D.; Granqvist, C.G. Critical Analysis of the Bennett–Riedel Attack on Secure Cryptographic Key Distributions via the Kirchhoff-Law–Johnson-Noise Scheme. *PLoS ONE*, 2013, 8, e81810, DOI: 10.1371/journal.pone.0081810. Available online: http://www.plosone.org/article/info%3Adoi%2F10.1371%2Fjournal.pone.0081810.
11. Kish, L.B.; Mingesz, R.; Gingl, Z.; Granqvist C.G. Spectra for the product of Gaussian noises. *Metrol. Meas. Syst.*, 2012, 19, 653–658, DOI: 10.2478/v10178-012-0057-0. Available online: http://www.degruyter.com/view/j/mms.2012.19.issue-4/v10178-012-0057-0/v10178-012-0057-0.xml.
12. Horvath, T.; Kish, L.B.; Scheuer, J. Effective privacy amplification for secure classical communications. *EPL*, 2011, 94, 28002, DOI:10.1209/0295-5075/94/28002. Available online: http://arxiv.org/abs/1101.4264.
13. Hao, F. Kish's key exchange scheme is insecure. *IEE Proc. Inform. Soc.*, 2006, 153, 141–142, DOI:10.1049/ip-ifs:20060068.
14. Kish, L.B. Response to Feng Hao's paper "Kish's key exchange scheme is insecure". *Fluct. Noise Lett.*, 2006, 6, C37, DOI: 10.1142/S021947750600363X. Available online: http://arxiv.org/abs/physics/0612193.
15. Smulko, J. *Fluct. Noise Lett.*, 2014, 13, 1450024, DOI: 10.1142/S0219477514500242.
16. Saez, Y.; Kish, L.B.; Mingesz, R.; Gingl, Z.; Granqvist, C.G. Current and voltage based bit errors and their combined mitigation for the Kirchhoff-law–Johnson-noise secure key exchange. *J. Comput. Electron.*, 2014, 13, 271–277, DOI 10.1007/s10825-013-0515-2. Available online: http://arxiv.org/abs/1309.2179.
17. Gunn L.J., Allison, A., Abbott, D. A directional wave measurement attack against the Kish key distribution system. Scientific Reports. 2014, 4, 6461, DOI 10.1038/srep06461.
18. Chen, H.P., Kish, L.B., Granqvist, C.G., Schmera, G. Do electromagnetic waves exist in a short cable at low frequencies? What does physics say? Fluctuation and Noise Letters 13 (2014) 1450016, DOI 10.1142/S0219477514500163. Available online: http://arxiv.org/abs/1404.4664.
19. Chen, H.P., Kish, L.B., Granqvist, C.G., Schmera, G. "On the "cracking" scheme in the paper "A directional coupler attack against the Kish key distribution system" by Gunn, Allison and Abbott". Metrology and Measurement Systems. 2014, 21, 389–400, DOI 10.2478/mms-2014-0033. Available online: http://arxiv.org/abs/1405.2034.